\input{aipcheck}

\documentclass[
  ,draft            
  ]
  {aipproc}

\layoutstyle{8x11single}

\newcommand{\cP}{{\cal P}}
\newcommand{\lnul}{{\ell_0}}

\renewcommand{\mathrm}{}
\newcommand{\myskip}[1]{}
\renewcommand{\mathit}{}

\newcommand{\lt}{{\ell_l}}
\newcommand{\Enl}{E_{nl}}
\newcommand{\El}{E_{nl}}

\newcommand{\cL}{{\cal L}}

\newcommand{\vL}{{\bf L}}
\newcommand{\cY}{{\cal Y}}

\newcommand{\vp}{{\bf p}}
\newcommand{\vq}{{\bf q}}

\newcommand{\vr}{{\bf r}}
\renewcommand{\vp}{{\bf p}}

\newcommand{\vn}{{\bf n}}

\newcommand{\p}{\partial}

\newcommand{\BEQ}{\begin{eqnarray}}
\newcommand{\EEQ}{\end{eqnarray}}
\newcommand{\BEA}{\begin{eqnarray}}
\newcommand{\EEA}{\end{eqnarray}}
\newcommand{\nn}{\nonumber }
\renewcommand{\d}{{{\rm d}}}

\newcommand{\om}{\omega}

\newcommand{\half}{\frac{1}{2}}

\newcommand{\azs}{{\alpha^2Z^2}}

\begin{document}

\title{Classical Phase Space Density for the Relativistic Hydrogen Atom}



\author{Th. M. Nieuwenhuizen}
{address={Institute for Theoretical Physics,
Valckenierstraat 65, 1018 XE Amsterdam, The Netherlands}}

\begin{abstract}
Quantum mechanics is considered to arise from an underlying classical structure
(``hidden variable theory'', ``sub-quantum mechanics''),
where quantum fluctuations follow from a physical noise mechanism.
The stability of the hydrogen ground state can then arise from a balance between
Lorentz damping and energy  absorption from the noise.
Since the damping is weak, the ground state phase space density  should predominantly be a function of
the conserved quantities, energy and angular momentum.

A candidate for this phase space density is constructed for ground state of the relativistic hydrogen problem
of a spinless particle.
The first excited states and their spherical harmonics are also considered in this framework.

The analytic expression of the ground state energy can be reproduced,
provided averages of certain products are replaced by products of averages.
This analysis puts forward that quantum mechanics may arise
from an underlying classical level as a slow variable theory,
where each new quantum operator relates to a new, well separated time interval.
\end{abstract}

\maketitle

\section{Introduction}

In this centenary of Einstein's Annus Mirabilis it is appropriate to reconsider
the foundations of physical theories. We shall, in particular, be interested in a
possible bridge between the three fields that Einstein pushed so much in 1905: brownian motion,
special relativity and quantum mechanics. Our focus is to investigate the possibility that quantum
mechanics arises as a statistical theory from a certain special relativistic theory,
with specific stochastic forces constituting the notorious ``quantum noise''.
Conceptually this will be a ``unified field theory'', which Einstein tried to formulate
in the second part of his life.
Without fully specifying this theory, we shall generalize squares of quantum wavefunctions
to phase space densities and see what it leads to.

In order to motivate our investigations, let us notice that the status of quantum mechanics
is still judged in two basically opposite ways.
The empirical view stresses that quantum mechanics, with its many postulates such as the superposition
principle, the uncertainty principle, the exclusion principle, the Born rule,
the collapse in measurements, is our best theory of nature,
proven to work in all cases where it was tested,
ranging from solid state physics and quantum chemistry
to high energy physics and the early Universe.
Most physicists even believe that quantum mechanics   is the ultimate theory of nature, that must be
unified with gravity, one popular route being string theory,
which adopts the additional postulates of supersymmetry and extra dimensions.

On a parallel track, there has always remained the fundamental question:
what is quantum mechanics   actually standing for?
Does nature have a reality: do particles exist as physical entities, possessing a position and a momentum;
is the moon there when nobody looks?
What is the physical mechanism behind the notorious ``quantum fluctuations'' and uncertainty relations?
What causes interference effects and the particle-wave duality?
Does each particle have its own wavefunction (standard Copenhagen interpretation),
what does a ``collapse of the wavefuction'' mean physically
\footnote{Nothing measurable according to Zbinden et al~\cite{zbinden}.} ?
For decades these questions have suffered from the lack of progress, in particular in
the quantum measurement problem, the only undisputed point of contact between quantum mechanics   and nature.
Very different interpretations have emerged~\cite{wh}: the Everett-Wheeler
relative state or multi-universe picture (collapse does not occur, but rather a branching);
the mind-body problem (the registration of the measurement in our mind or a machine
matters) (Wigner, see ~\cite{wh}), the wavefunction codes our state of knowledge~\cite{Mermin}
or our state of belief~\cite{Fuchs}.

At present a variety of different approaches support the possibility that at the quantum level,
nature has a `natural' (i.e. more classical than long supposed) behavior. We may mention:
arguments against the conclusion that nature should be non-local because of violations of Bell inequalities
~\cite{HessPhillip, Volovich, Willeboordse}; the demonstration that quantum probabilities including
interference effects are compatible with classical probability theory~\cite{Khrennikov};
the long effort of Stochastic Electrodynamics~\cite{delaPenaCettoBook} and its later
version called Linear  Stochastic Electrodynamics~\cite{delaPenaCettoBook,delaPenaCetto};
the demonstration of entanglement in classical brownian motion~\cite{AKNbrownentanglement};
the description of an electron with its spin as a soliton (over-extremal Kerr-Newman black hole)
in electro-gravity~\cite{Carter,ArcosPereira};
the many arguments in favor of the statistical interpretation
~\cite{ballentine,vKampen,BalianAmJP,demuynck,ABNBE,ABNmeasEPL};
explicit solutions for the quantum measurement problem~\cite{ABNBE,ABNmeasEPL}.

Let us go in some detail about the measurement process.
By taking into account an explicit model for the apparatus, the quantum measurement
problem was recently solved in our group in two situations: in the first the position of a boson is
measured by an apparatus that consists of many bosons. The apparatus starts close to an (ideal)
Bose-Einstein phase transition and is driven into it by the measurement~\cite{ABNBE}.
In the second model the $z$-component of a spin-$\half$ is measured by an apparatus consisting
of many spins-$\half$~\cite{ABNmeasEPL} that start out as a paramagnet and are driven into the up or
down ferromagnetic phase, according to the sign of the measured spin component.
In both cases the apparatus also consists of a heat bath, which originates
from other degrees of freedom, in the magnetic case the lattice vibrations.

Quantum mechanics  being a probabilistic theory, it should come as no surprise that what could
be calculated was the statistical distribution of outcomes for an ensemble of measurements
on an ensemble of identically prepared systems or identical preparations
of a single system.
Technically this comes in because the whole exercise is to determine the
post-measurement density matrix given its pre-measurement form.
The process goes in two steps: when coupling the system to the apparatus, there
appears a very fast collapse of the off-diagonal terms (the death of Schr\"odinger cats).
 This ``collapse of the wavepacket'' is a unitary evolution in the large Hilbert space of system
and pointer degrees of freedom of the apparatus (thus not involving the bath), very
similar to the process of dephasing in NMR physics. The collapse time depends as an inverse power
on the number of particles in the apparatus, so it is strongly microscopic when the apparatus is macroscopic.
On a later stage the effect of the bath is to make this collapse definite by decoherence.
Schr\"odinger cat terms thus disappear automatically, because of the macroscopic size
of the apparatus. The second step of the measurement is its registration,
which involves the diagonal elements of the density matrix and appears to have classical
features~\cite{ABNNapels}.
Indeed, for the magnetic model, the solution of this part
of the problem is just what one might have anticipated from purely classical thermodynamic
reasoning~\cite{ABNNapels}. In particular, a bath is needed simply because
energy has to be dumped when the macroscopic pointer variable (the magnetization of the apparatus)
goes from its initial metastable paramagnetic value to its up or down ferromagnetic value.
Also in cloud chambers, the energy released upon creating bubbles in an oversaturate
liquid has to be transfered to the environment and this very mechanism allows the selection
of circumstances where these bubbles have a desired average size.

From the recent solutions of two realistic models for the quantum measurement process,
it became clear that the statistical interpretation of quantum mechanics
 allows a simple interpretation of this process during and after the
measurement. To mention one problem that is immediately solved in this way:
the fundamental question how quantum mechanics  can describe collapse of the measured operator
has the answer: quantum mechanics  describes only the
statistics of those collapses; it is not the proper theory to describe an individual collapse.
Spontaneous collapse models (GRWP theories), that try to treat individual collapses from a slightly
extended quantum mechanics, attempt to heal a problem that has its roots at a more fundamental level.
Indeed, for a two slit experiment quantum mechanics does not describe any individual spot
on the photographic plate, but only the density of many spots.

Thus, we conclude that {\it Quantum mechanics is not a theory of nature itself, but a theory for the
statistics of outcomes of experiments}. This situation is very similar to the one of thermodynamics,
where the underlying level of, say a Lennard-Jones gas, is Newton's equations of motion.
The main difference is, however, that experiments have to be made with apparati which themselves
are always macroscopic and therefore influence the microscopic dynamics.
Since it is the task of physicists to describe nature, the search for a sub-quantum
theory (hidden-variables theory) is, in our eyes, completely justified.
The program to search this may contain the following steps

1) For point particles, show compatibility between classical phase space density structures and
quantum structures.

2) Find the underlying stochastic theory and isolate the stochastic forces which cause
the ``quantum fluctuations''. Show compatibility with quantum mechanics.

3) Find soliton-like solutions for elementary particles such as the electron.
An interesting attempt sees an electron as a Kerr-Newmann black hole, that is,
as a charged ring with diameter given by the Compton wavelength~\cite{Carter, ArcosPereira}.
On the ring a current runs with the speed of light, causing a spin, with $g=2$,
as in the Dirac theory for the electron.
In these theories the parameters must be fixed to the physical values by adding
terms to the Einstein-Maxwell theory. A possibility is to couple
torsion with the electromagnetic field tensor~\cite{Nh2bpub}, which might
fix charge and/or spin, though not yet the mass. Within this framework the Pauli principle
might arise as an energetic constraint on the behavior in the far field region, and
Einstein photons as massless spin-one solitons. Let us notice that indistinguishably is
a natural aspect of solitons that wind up in a chaotic motion.

4) Show that the consistency still works for the ``stochastic soliton mechanics''
of solitons in the random electromagnetic background, with ``photons'' appearing as
classical EM waves and as ``Einstein'' solitons.

Clearly, the program may run into contradictions at many points:
apart from internal consistency, compatibility with a huge amount of
experiments is required. But if Einstein was right after all, then there should be some road to the
underlying structure, and this one seems to us the simplest, so it deserves to be investigated further.

In this work we shall restrict ourselves to point 1) of the program and observe that already
this one is more complicated than sketched here.

\section{Motivation from Stochastic Electrodynamics}

Quantum mechanics  has both a particle and wave nature. In a subquantum approach, it is clear how to
think of the particles, to start they are just considered as points,
so the next question: what are the waves made of.
In the simplest approach, one assumes that the waves are due to fields that we know already.
The first candidate then is Electromagnetism. One has then arrived at the so-called
Stochastic Electrodynamics (SED), where the zero-point fluctuations known as a slogan in
quantum mechanics  are attributed to a physical entity, namely classical random EM fields
~\cite{delaPenaCettoBook}.
To get the proper connection, one then has to choose
the energy content of each mode as $\half\hbar\omega$, in which the prefactor $\hbar$
is just a system parameter, chosen to coincide with Planck's constant. The related
spectral density is just the zero point part of the Planck spectrum,
 $\rho(\omega)=\hbar\omega^3/(2\pi^2 c^3)$.
This $\om^3$-dependence is precisely the one consistent with Lorentz invariance and
with thermodynamics.

The theory of SED has a long history of both success and failures. SED has explained
the harmonic oscillator and, with it, electric dipole harmonic oscillators,
the Casimir and Casimir/Polder effects, and the Unruh effect.

Away from harmonic systems, SED has a long record of failures.
First, a connection with the Schr\"odinger equation has never been found.
Second, the role for excited states has been discussed only in the margin.
It should be admitted that, whereas Schr\"odinger dynamics at first neglects the
decay of excited states, the structure of SED is that of
time-dependent perturbation theory, where the full time-dependence is taken into account
from the start.
Third, a long list of non-linear problems has been studied, where results
different from quantum mechanics  emerge, see e.g. \cite{Pope}.

The most obvious example will be  the subject of this study: the ground state of the
hydrogen problem. This is a rather clean problem that should be understood before
conclusions are drawn about validity or invalidity and about multi-level atoms.
Indeed, in the hydrogen problem no internal structure should be expected that complicates
the orbiting of the point-electron in the central force field of the infinitely heavy nucleus.
Theoretical results so far are meager, predicting stability radius of the orbit equal
to infinity~\cite{delaPenaCettoBook}. Surprisingly, a numerical analysis
by Zhou and Cole supports an SED explanation of the quantum ground  state density~\cite{Cole}.

That the hydrogen problem should have a stable solution within SED, was put forward
long ago: when the electron is far from the  nucleus, its orbital frequency is low
and so should be the relevant frequencies of the stochastic force. Due to the $\om^3$ spectrum,
there should essentially be no random force at all. What remains is the Lorentz damping that
brings the electron closer to the nucleus. For small distances, on the other hand, the relevant
frequencies should be large, and the spectrum predicts a strong kicking due to the random
forces, preventing electron to fall into the nucleus due to Lorentz damping alone.
There thus appears a stable atomic state. Is this  the one we know from quantum mechanics?
So far attempts to show this have failed, including those by ourselves~\cite{Nh2bpub}.
For this reason it would be interesting to corroborate the numerical analysis of~\cite{Cole}.

\renewcommand{\thesection}{\arabic{section}}
\section{Quantum mechanics of the relativistic hydrogen atom}
\setcounter{section}{1}
\setcounter{equation}{0}
\setcounter{figure}{0} \renewcommand{\thesection}{\arabic{section}.}

The Hamiltonian for a spinless boson with rest mass $m$ and charge $e$,
 in the presence a central charge $-Ze$, reads
\BEQ H=\sqrt{m^2c^4+\vp^2c^2\,}-\frac{Ze^2}{r}
=\gamma mc^2-\frac{Ze^2}{r}\EEQ

We express $\vr$, $\vp$ and time in atomic units,
\BEQ
\frac{\vr}{a_0}\to \vr,\quad \frac{t}{\tau_0}\to t,\qquad
\frac{\vp\,\tau_0}{ma_0} \to \vp \EEQ
with Bohr radius and Bohr time
\BEQ a_0=\frac{\hbar}{\alpha Zmc}=\frac{1}{Z}\,5.29\,10^{-11}m,\qquad
\tau_0=\frac{\hbar}{\alpha^2Z^2mc^2}=\frac{1}{Z^2}\,2.418 \,\,10^{-17}s \EEQ
with numerical values for the electron. This scaling also implies
\BEQ \frac{E}{mc^2}\to E, \quad \frac{L}{\hbar}\to L,\quad \EEQ
and brings the scaled form
\BEQ H=\sqrt{1+\azs\vp^2}-\frac{\azs}{r}
=\gamma -\frac{\azs}{r}\EEQ

Let us denote quantum operators by a subscript $op$.
The Schr\"odinger equation reads

\BEQ (H_{op}-E)|\psi\rangle=(\,\sqrt{1+\azs \vp^2_{op}}-\frac{\azs}{r_{op}}-E)|\psi\rangle=0
\EEQ
this is equivalent to
\BEQ
(\half\vp^2_{op}-\frac{\azs}{2r^2_{op}}-\frac{E}{r_{op}})|\psi\rangle
=-\frac{1-E^2}{2\azs}|\psi\rangle
\EEQ
In the position basis this reads
\BEQ
(-\half \p_r^2-\frac{1}{r}\p_r +\frac{L_{op}^2-\azs}{2r^2}-\frac{E}{r}+\frac{E^2}{R})\psi(r)=0,
\qquad \frac{1}{R}=\frac{1-E^2}{2\azs\,E^2}
\EEQ
When scaling $r\to \tilde r/E$, $\vp\to E\tilde\vp$, and using that $L_{op}^2$ has
eigenvalues $l(l+1)$ with non-negative integer $l$, this  becomes
\BEQ  \label{scaledSE}
(-\half\p_{\tilde r}^2-\frac{1}{\tilde r}\p_{\tilde r} +\frac{l(l+1)-\azs}{2\tilde r^2}
-\frac{1}{\tilde r})
\psi=-\frac{1}{R}\psi
\EEQ
so $-1/R$ is the associated non-relativistic energy.
It is customary define the effective angular momentum value $\ell_l$ by
\BEQ\lt(\lt+1)=l(l+1)-\azs,\qquad  \lt=-\half+\half\sqrt{(2l+1)^2-4\azs}
\approx l-\frac{\azs}{2l+1}
\EEQ
Now the whole problem can be solved by analogy with the non-relativistic situation.
The ground state wavefunction is
\BEQ R_{10}=Cr^{\ell_0}\,e^{-rE_{10}/(\ell_0+1)},\qquad
\EEQ

The so called Yrast states have $(n,l)=(n,n-1)$ and wavefunctions
\BEQ R_{nl}=Cr^{\lt}e^{-r\Enl /(\lt+1)} \EEQ
They have energy
\BEQ \label{Enl=} E_{n,l}=\{1+\frac{\azs}{[n-l-\half+\sqrt{(l+\half)^2-\azs}]^2}\}^{-1/2}
\approx 1-\frac{\azs}{2n^2} -\alpha^4Z^4[\,\frac{1}{n^3(2l+1)}-\frac{3}{8n^4}\,]
\EEQ
a formula which is valid for all levels, that is to say, for $n=1,2,\cdots$ and $0\le l\le n-1$.

Let us mention that when the spin of the electron is taken into account, one has to
replace $l\to j=l\pm\half$.

\renewcommand{\thesection}{\arabic{section}}
\section{statistical mechanics of the relativistic Coulomb problem}
\setcounter{equation}{0}\setcounter{figure}{0} \setcounter{section}{2}
\renewcommand{\thesection}{\arabic{section}.}

The Schr\"odinger equation works so well for the hydrogen atom, because corrections arising from the coupling to the
electromagnetic field are weak. They lead to the Lamb shift, which is of order $\alpha^5Z^4\ln\alpha Z$ in
the present units. So it is weaker than the fine structure, represented by the $\alpha^4Z^4$ term in (\ref{Enl=}).
The very same mechanism makes the lifetimes of excited states much larger than the Bohr time.

To focus our line of reasoning, let us consider the theory of stochastic electrodynamics (SED). Here random
electromagnetic fields are assumed to exist.
In this theory a Fokker-Planck approach may be formulated for the statistics of the random orbits
($\underline{\vr}(t),\underline{\vp}(t)$)
~\cite{delaPenaCettoBook,Nh2bpub}. The phase space density will satisfy an equation of the form
\BEQ P(\vr,\vp,t)=\langle\delta(\underline{\vr}(t)-\vr)\delta(\underline{\vp}(t)-\vp)\rangle \EEQ
will satisfy
\BEQ \p_tP=-{\cal L}P+{\it Lorentz\,\, damping}+{\it di\!f\!\!f\!usion\,\, terms}\EEQ
with drift given by the Liouvillian
\BEQ \cL\equiv \frac{1}{\gamma}\vp \cdot\nabla_\vr -\frac{\vr}{r^3}\cdot \nabla_\vp
\EEQ
Since damping and diffusion terms are of order $\alpha^3$, the stationary distribution should, to this order,
 be a function of the conserved quantities, angular momentum $L$ and energy $E$. This remains valid if
one goes beyond the Fokker-Planck equation and for other stochastic theories.
So the weakness of the Lamb shift finally allows us to consider equilibrium structures of a class
of stochastic theories, without going to specific details.

The ground state should follow from the dynamical problem
$H\equiv\gamma-\azs/r=E$ or, equivalently,
\BEQ\label{Egam=}
\half \vp^2-\frac{\azs}{2r^2}-\frac{E}{r}=\frac{E^2-1}{2\azs}
\EEQ
There will also be effects of damping and noise, but they enter here only at order $\alpha^3$,
and should explain the Lamb shift. We shall neglect those effects in the present paper.
If we define $R$ by
\BEQ\label{ER=}
E=\sqrt{1+\frac{\alpha^4Z^4}{R^2}}-\frac{\azs}{R}=\left(1+\frac{2\azs }{RE}\right)^{-1/2}
\EEQ
the dynamical problem becomes
\BEQ \half \vp^2-\frac{\azs}{2r^2}-\frac{E}{r}=-\frac{E}{R}
\EEQ

Close to $r=0$ and for $p_r=0$ it is seen that $p_\perp=\alpha Z/r$,
implying that $L\ge \alpha Z$ and $R\ge r$.
Let us denote
\BEQ\om=\sqrt{1-\frac{Z^2\alpha^2}{L^2}},\qquad \om^2L^2=L^2-\azs\EEQ

We shall take spherical coordinates in $\vr$ space.
In the frame along $\vr=r(0,0,1)$ we introduce   for $\vp$
the angles $\mu$ and $\nu$ with $0\le\mu\le\pi$ and $0\le\nu\le2\pi$,
\BEQ
\vp=(p_{\perp1},p_{\perp2},p_r)=
p(\sin\mu\cos\nu,\sin\mu\sin\nu,\cos\mu).
\EEQ

In special relativity the invariant volume elements in coordinate and momentum space are
\BEQ \d V_r=\d^3r=\d r \d\theta\d\phi r^2\sin\theta,\qquad \d V_p=\frac{\d^3p}{\gamma}\EEQ
We have, using  that $L\d L=\om L\d(\om L)$,
\BEQ \d V_p=\frac{\d p_r\d \nu\,\d p_\perp p_\perp}{\gamma}=\frac{\d p_r\d\nu L\d L}{r^2\gamma}
=\frac{\d p_r\d\nu \,\om L\d (\om L)}{r^2\gamma}\EEQ
We go to the new variables $\mu$ and $R$ by the transformation
\BEQ
\om L=r\sqrt{2E(\frac{1}{r}-\frac{1}{R})}\,\,\sin\mu,\qquad
p_r=\sqrt{2E(\frac{1}{r}-\frac{1}{R})}\,\cos\mu
\EEQ
and have
\BEQ \d p_r\d(\om L)=r\d\mu\d R\,\frac{\p}{\p R}(\frac{E}{r}-\frac{E}{R})
=\frac{r\d\mu\d R\,\,\gamma E}{R^2\sqrt{1+\alpha^4Z^4/R^2}},\qquad \gamma=E+\frac{\azs}{r}
\EEQ
This implies the volume element in momentum space
\BEQ \d V_p=
\frac{\d\mu\d\nu\d R}{2R^2\Phi(E)}
\sqrt{\frac{2}{Er}-\frac{2}{ER}}\,\sin\mu\,\qquad
\Phi(E)=\frac{\sqrt{1+\alpha^4Z^4/R^2}}{2E^2}
\EEQ
Notice that the $r$-dependence through $\gamma$ has dropped out and that $\Phi=\half$ non-relativistically.
If the phase space density has the form $\cP(E,L)=\,\om LR^3\Phi(E)\exp(-aR)$, then we have
\BEQ \d V_p\,\cP=\d\mu\d\nu\d R \,(R-r)e^{-aR}\sin^2\mu \EEQ
The momentum integrals, that is to say, the integrals over $\mu$, $\nu$ and $R\ge r$, then yield
\BEQ  \int_\vp\d V_p\,\cP=\frac{\pi^2}{a^2}e^{-ar} \EEQ
which for $a=2$ is just the square of the non-relativistic ground state wavefunction.
Furthermore, since
\BEQ \frac{\om^2L^2R}{2E}=r(R-r)\sin^2\mu,\EEQ
powers of $r$ in the squared wave functions can be generated by powers of this ratio in the
phase space density.
So we may choose the following combination of the conserved quantities $E$, $R(E)$ and $L$
for ground state $(1,0)$ and, more generally, the Yrast states $(n,l)=(n,n-1)$ with $n=1,2,\cdots$
\BEQ \label{Pc=}
\cP(R,L)=C\om LR^3\Phi(E)\left(\frac{\om^2L^2R}{2E}\right)^{2\lt}
\,e^{-2\,R\,\El /(\lt+1)}
\EEQ
Fixing $C$ this leads to our final proposition
\BEQ\label{density=} \d V_p\cP=\d R\,\frac{\d\mu}{\pi}\,\frac{\d\nu}{2\pi}\,\,
\frac{2^{6 + 8\lt}\El^{5 + 4 \lt}}{(1 + \lt)^{6+ 4\lt}\Gamma(3 + 4 \lt)}
r^{2\lt}(R-r)^{1 + 2\lt}(\sin\mu)^{2 + 4 \lt}e^{-2\,R \El /{(1 + \lt)}}
\EEQ
This form is very appealing, though perhaps not unique.
The angular integrals yield
\BEQ
\int_{\mu,\nu}\d V_p\cP=
\frac{\d R\,\,2^{4 + 4\lt}\El^{5 + 4 \lt}}{(1 + \lt)^{6+4 \lt}\Gamma^2(2 + 2\lt)}
 r^{ 2\lt}(R-r)^{1 + 2 \lt}e^{-2 R\El /{(1 + \lt)}}
\EEQ
and the full momentum average brings the proper squares of the wavefunctions,
\BEQ
\int_\vp\d V_p\cP=\int_{R,\mu,\nu}\d V_p\cP=
\frac{2^{2 + 2\lt}\El^{3 + 2 \lt}}{(1 + \lt)^{4 + 2\lt}\Gamma(2 + 2\lt)}
r^{ 2\lt}\, e^{-2r \El /{(1 + \lt)}}
\EEQ
On the other hand, the spatial average yields
\BEQ \d V_p\int_0^\infty\d r\,r^2\cP=\d R\,\frac{\d\mu}{\pi}\,\frac{\d\nu}{2\pi}\,\,
\frac{\El^{5 + 4 \lt}}{(1 + \lt)^{6 + 4 \lt}
\Gamma(\frac{3}{2} + 2\lt)\Gamma(\frac{5}{2} + 2\lt)}
R^{4 + 4\lt}\sin^{2+4\lt}\mu\,  e^{-2 R\El /{(1 + \lt)}}
\EEQ
The angular integral brings
\BEQ
\int_0^\infty\d r\,r^2\int_{\mu,\,\nu}\d V_p\cP=\d R\,
\frac{2^{3 + 4\lt}\El^{5 + 4 \lt}}{(1 + \lt)^{6 + 4 \lt}\Gamma(4 + 4\lt)}
R^{4 + 4\lt}  e^{-2 R\El /{(1 + \lt)}}
\EEQ
This implies the expectation values
\BEQ &&\langle\frac{1}{r}\rangle= \frac{\El}{(1 + \lt)^2}=\langle\frac{2}{R}\rangle,\qquad
\langle\frac{1}{r^2}\rangle= \frac{2\El^2}{(1 + \lt)^3(1+2\lt)},
\qquad
\langle\frac{1}{R^2}\rangle=\frac{\El^2}{(1 + \lt)^3(3 + 4\lt)}
\EEQ
Eq. (\ref{ER=}) then implies, to order $\alpha^4$, the classical expectation value
\BEQ
\label{Ewrong=}
\langle E\rangle_{nl;\,\,cl} =1-\azs\langle\frac{1}{R}\rangle+ \half\alpha^4Z^4\langle\frac{1}{R^2}\rangle
=1-\frac{\azs}{2}-\frac{(4n^2+4n-1)\alpha^4Z^4}{4n^4(8n^2-6n+1)}
=\El+\frac{\alpha^4Z^4}{8 n^4(4n-1)}\EEQ
This is not the quantum  mechanical result, so the approach fails at order $\alpha^4$,
that is to say, for the fine structure.

\renewcommand{\thesection}{\arabic{section}}
\section{Neglecting correlations}
\setcounter{equation}{0}\setcounter{figure}{0} \setcounter{section}{3}
\renewcommand{\thesection}{\arabic{section}.}
It is easily verified that the proper $\alpha^4$ term of the quantum eigenenergies (\ref{Enl=})
emerges is we replace the term $\langle R^{-2}\rangle$ of Eq. (\ref{Ewrong=})
by $\langle R^{-1}\rangle^2$, thus neglecting the correlator
$\langle R^{-2}\rangle-\langle R^{-1}\rangle^2$.
This approach appears to work  to infinite order.

\subsection{Yrast states}

Surprisingly, the quantum mechanical result does emerge if,
after expanding the square root (\ref{ER=}) in powers of $\alpha$, we replace
in the successive terms the averages of products by products of averages.
The result can then be resummed,
\BEQ \label{E0SED}\langle\langle E\rangle\rangle_{nl;\,cl} \equiv
\sqrt{1+\alpha^4Z^4\langle\frac{1}{R}\rangle^2}-\azs\langle\frac{1}{R}\rangle
=\sqrt{1+\frac{\alpha^4Z^4\El^2}{4(1+\lt)^4}}-\frac{\azs \El}{2(1+\lt)^2}
\EEQ
From the expression (\ref{Enl=}) one may derive the relation
\BEQ \label{azs=}
\azs=\frac{1-\El^2}{\El^2}(n-l+\lt)^2
\EEQ
For the Yrast states we then observe agreement with the quantum eigenvalues for the energy,
\BEQ \langle\langle E\rangle\rangle_{nl;\,cl}=
\sqrt{\frac{(1+\El^2)^2}{4\El^2}}-\frac{1-\El^2}{2\El}
=\El\EEQ
 to all orders in $\alpha$. Physically such infinite precision of fortuitous, since
there are Lamb shift  corrections of order $\alpha^5$.

By working with expression (\ref{Egam=}), we have another approach, which acts as
a consistency test. If we separate the averages and calculate the average of
$E$ as above, we find
\BEQ \langle\langle p_r^2\rangle\rangle=
\langle\langle2E(\frac{1}{r}-\frac{1}{R})\,\,\cos^2\mu \rangle\rangle=
2\langle\langle E\rangle\rangle\,\,\langle(\frac{1}{r}-\frac{1}{R})\,\,\cos^2\mu \rangle\,
=\frac{\El^2}{4(1 + \lt)^3}\EEQ
and
\BEQ
\langle \frac{\om^2L^2}{r^2}\rangle
=\frac{\El^2(3 + 4\lt)}{4(1 + \lt)^3}.
\EEQ
These results bring
\BEQ\label{davL2}
 \langle\langle\frac{L^2}{r^2}\rangle\rangle=
\langle\frac{\om^2L^2}{r^2}\rangle+\langle\langle\frac{\azs}{r^2}\rangle\rangle
=\langle \frac{\om^2L^2}{r^2}\rangle+\azs\langle\frac{1}{r}\rangle^2
=\frac{1}{(1+\lt)^2}-\frac{\El^2}{4(1+\lt)^3}\EEQ

Combining both results brings simply
\BEQ \langle\langle\vp^2\rangle\rangle=\langle\langle p_r^2\rangle\rangle
+\langle\langle\frac{L^2}{r^2}\rangle\rangle
=\frac{1}{(1+\lt)^2}
\EEQ
When averaging Eq. (\ref{Egam=}), we again expand in powers of $\azs$ and
replace repeatedly averages of products by products of averages.
We end up with the alternative expression
\BEQ \langle\langle E\rangle\rangle_{nl;\,cl}'=
\sqrt{1+\azs\langle\langle\vp^2\rangle\rangle}-\langle\langle \frac{\azs}{r}\rangle\rangle=
\sqrt{1+\frac{\azs}{(1+\lt)^2}}-\frac{\azs \El}{(1+\lt)^2}
\EEQ
The leading correction of this square root
is stronger than the one of order $\alpha^4Z^4$ of the expression (\ref{E0SED}).
But by using eq. (\ref{azs=}) is easy to see that
\BEQ
\langle\langle E\rangle\rangle_{nl;\,cl}'=
\sqrt{\frac{1}{\El^2}}-\frac{1-\El^2}{\El}=\El
\EEQ
 confirming consistency of the approach, provided it includes the replacement
$\langle\langle r^{-2}\rangle\rangle
\to\langle r^{-1}\rangle^2$ done in Eq. (\ref{davL2}).

\subsection{The $2s$  state}
The Yrast states could be special, because their spatial wavefunctions consist of a
single term.
We now investigate a non-Yrast state, the $2s$ state, which has and additional polynomial in its
wavefunction:
\BEQ R_{20}=\left\{\frac{ (1 + \lnul) 2^{3 + 2\lnul}}
{(2 + \lnul) (4 + 3\lnul)^{3/2+ \lnul}\Gamma(2 + 2\lnul)}\right\}^{1/2}
\, r^{\lnul}
\left(1 - \frac{r}{(1 + \lnul)\sqrt{4 + 3\lnul}}\right) \exp\left(- \frac{r}{\sqrt {4 + 3\lnul}}\right)
\EEQ
The function $R_{20}(\frac{r}{E_{20}})$ is orthogonal to
$R_{10}(\frac{r}{E_{10}})$, as it should in view of the scaled Schr\"odinger equation (\ref{scaledSE}).
The probability density $P_{20}(r)=R_{20}^2(r)$  may derive from a phase space density,
\BEQ&&
\cP_{20}(E,L,R(E))=\frac{2^{7 + 8 \lnul}\,\,}
{(2 + \lnul)\pi^2\Gamma(6+4 \lnul)(4 + 3 \lnul)^{9/2+2\lnul}}\,\,
\om LR^3\Phi(E)\left(\frac{\om^2L^2R}{2E}\right)^{2\lnul}
\exp\left(-\frac{2R}{\sqrt{4 + 3\lnul}} \right)\nn \\ &&\times
\left[(1 + \lnul)^2 (4 + 3 \lnul)^2 (3 + 4 \lnul)(5 + 4 \lnul) - 8 (1 + \lnul)(4 + 3 \lnul)(5 + 4 \lnul)
\frac{\om^2L^2R}{2E} + 16 \left(\frac{\om^2L^2R}{2E}\right)^2\,\,\right]
\EEQ
which after integrating over the angles $\mu$ and $\nu$ reads
\BEQ
\int_{\mu,\nu}\d V_p\cP_{20}&=&
\,\frac{\d R\,\,2^{8 + 4\lnul} }
{(4 + 3\lnul)^{9/2 + 2\lnul}\Gamma(3+2\lnul)\Gamma(5+2\lnul)}\,
r^{2\lnul}(R-r)^{1 + 2\lnul}\exp\left(-\frac{2R}{\sqrt{4 + 3\lnul}}\right) \\
&\times& [(1 + \lnul )^3(3 + 2\lnul )(4 + 3\lnul)^2
- 2(1 + \lnul )(3 + 2 \lnul )(4 + 3\lnul )r(R-r) + 2r^2(R-r)^2] \nn
\EEQ
and next
\BEQ \int_0^R\d r\, r^2\int_{\mu,\nu}\d V_p\cP_{20}
&=&\frac{2^{9 + 4 \lnul }(3 + 2 \lnul )}{(4 + 3 \lnul )^{9/2+2\lnul}\Gamma(9 + 4 \lnul )} R^{4 + 4 \lnul }\,
\exp\left(-\frac{2R}{\sqrt{4 + 3\lnul}}\right)
\\& \times&
[(1 + \lnul )^2(4 + 3 \lnul )^2(5 + 4 \lnul )(7 + 4\ \lnul )
- 2(1 + \lnul )(4 + 3 \lnul )(7 + 4 \lnul ) R^2 + R^4\,] \nn
\EEQ
This  brings the averages
\BEQ\langle\frac{1}{r}\rangle=\frac{1}{(2 + \lnul)\sqrt{4 + 3\lnul}},\qquad
  \langle\frac{1}{R}\rangle=\frac{1}{2(2 + \lnul)\sqrt{4 + 3\lnul}},\EEQ
which is just what is needed, because we get using $\azs=-\lnul(\lnul+1)$,
\BEQ \langle\langle E \rangle\rangle_{20;\,cl}=
\sqrt{1+\alpha^4Z^4\langle\frac{1}{R}\rangle^{^2}}-\langle\frac{\azs}{R}\rangle
=\frac{8+7\lnul+\lnul^2}{2(2 + \lnul)\sqrt{4 + 3\lnul}}+\frac{\lnul(\lnul+1)}
{2(2 + \lnul)\sqrt{4 + 3\lnul}}=\frac{2+\lnul}{\sqrt{4 + 3\lnul}}=E_{20}\EEQ

We can further verify that
\BEQ
 \langle\langle \vp^2 \rangle\rangle\equiv2 \langle\langle E \rangle\rangle\,[
 \langle \frac{1}{r} \rangle- \langle \frac{1}{R}\rangle\,]\,
+\azs \langle \frac{1}{r} \rangle^2=\frac{1}{(2+\lnul)^2}
\EEQ
implying that also the alternative calculation leads to the quantum mechanics   result,
\BEQ
 \langle\langle E \rangle\rangle'_{20;\,cl}=\sqrt{1+\azs  \langle\langle\vp^2} \rangle\rangle
-\azs \langle \frac{1}{r} \rangle=
\frac{\sqrt{4+3\lnul}}{2+\lnul}+
\frac{\lnul(\lnul+1)}{({2+\lnul})\sqrt{4+3\lnul}}
=\frac{2+\lnul}{\sqrt{4+3\lnul}}
\EEQ
It appears that the approach is non-unique. The density $R_{20}^2(r)$ can also be obtained from
\BEQ&&
\cP_{20}'=\frac{2^{6 + 4 \lnul}(1+\lnul)\,\,}
{(2 + \lnul)\Gamma^2(3+2 \lnul)(4 + 3 \lnul)^{7/2+2\lnul}}\,\,
\om LR^3\Phi(E)\left(\frac{\om^2L^2R}{2E}\right)^{2\lnul}
\exp\left(-\frac{2R}{\sqrt{4 + 3\lnul}} \right)
\nn\\&&\times
\left[(1+\lnul)(4+3\lnul)(7+8\lnul)-8(1+\lnul)\sqrt{4+3\lnul}\,R+2R^2\,\right]\,
\left[2\pi B(\frac{3}{2}+2\lnul,\half)\right]^{-1}
\EEQ
This form has the same  expectation for $\langle1/r\rangle$ and  $\langle1/R\rangle$, and thus for energy.
The two forms can thus be mixed, which leads to a one-parameter family of solutions;
the proper choice cannot be fixed in the present approach.
By viewing such mixings as a way to fix the $r^2$ term of the square of the polynomial
of $R_{20}$, one might wonder whether the $r$ term can be changed with a second independent
parameter. It has been checked, however, that that would modify $\langle1/R\rangle$.

\renewcommand{\thesection}{\arabic{section}}
\section{Phase space forms for squares of spherical harmonics}
\setcounter{equation}{0}\setcounter{figure}{0} \setcounter{section}{4}
\renewcommand{\thesection}{\arabic{section}.}
Previous expressions $R_{n0}^2$ for $s=0$ states are still multiplied by a factor
$Y_{00}^2=1/(4\pi)$, which drops out after the integration over the angles $\theta$ and $\phi$
of the coordinate vector. Let us denote the generalization of $Y_{lm}^\ast Y_{l'm'}$ by
$\cY_{lm}^{l'm'}$. Then we can likewise multiply our results for $\cP_{10}$ and $\cP_{20}$
by $\cY_{00}^{00}={1}/({4\pi})$ and cancel it by the angular integrals.

For the $\cP_{21}$ distribution we only considered the radial part;
there is still the question how to deal with the factor $Y_{1m}^2$
and the same question arises for the higher $l$ states at $n\ge 2$.
According to the quantum solution, such diagonal terms should have no
time-dependence, so also $Y_{1m}^2$ is be generalized to a function depending
on conserved quantities only.

Let us notice that in the frame along $\vr$ we can introduce angles
$\mu$ and $\nu$ with $0\le\mu\le\pi$ and $0\le\nu\le2\pi$,
\BEQ \vr=r(0,0,1),\qquad \vp=p(\sin\mu\cos\nu,\sin\mu\sin\nu,\cos\mu)=
(p_{\perp1},p_{\perp2},p_r),\qquad
\vL=pr\sin\mu\,(-\sin\nu,\cos\nu,0) \EEQ
In the laboratory frame this implies
\BEQ && \vr=r(\sin\theta\cos\phi,\sin\theta\sin\phi,\cos\theta),\qquad
\vL=pr\sin\mu\,\hat\vL,\qquad \nn\\&&
\hat\vL=(-\cos\theta\cos\phi\sin\nu-\sin\phi\cos\nu,
-\cos\theta\sin\phi\sin\nu+\cos\phi\cos\nu,\sin\theta\sin\nu)
\EEQ
Specifying the $z$-axis as our preferred axis, we have $\hat L_z=\sin\theta\sin\nu$
and we define
\BEQ
\cY_{1-1}^{1-1}=\frac{3}{4\pi}(\hat L_z^2- \hat L_z),\qquad
\cY_{10}^{10}=\frac{3}{4\pi}(1-2\hat L_z^2),\qquad \cY_{11}^{11}=\frac{3}{4\pi}(\hat L_z^2+\hat L_z).
\EEQ
In general with $\vn$ a fixed unit vector this would read
\BEQ \label{Lzlinear}
\cY_{1-1}^{1-1}=\frac{3}{4\pi}\left[(\vn\cdot\hat \vL)^2- \vn\cdot\hat\vL\,\right],\qquad
\cY_{10}^{10}=\frac{3}{4\pi}\left[1-2(\vn\cdot\hat \vL)^2\right], \qquad
\cY_{11}^{11}=\frac{3}{4\pi}\left[(\vn\cdot\hat \vL)^2+ \vn\cdot\hat \vL\right],\qquad\EEQ
The odd terms are chosen linear, as this seems most natural to us.
To settle their form uniquely is, however, not possible in the present
approach, since for that aim the underlying physics should be specified further.

The $\vp$-integration produces an integral over $\nu$, which brings the
desired results
\BEQ
\int_0^{2\pi}\frac{\d\nu}{2\pi} \cY_{11}^{11}=
\int_0^{2\pi}\frac{\d\nu}{2\pi} \cY_{1-1}^{1-1}=\frac{3}{8\pi}\sin^2\theta=|Y_{1\pm1}|^2,\qquad
\int_0^{2\pi}\frac{\d\nu}{2\pi} \cY_{10}^{10}=\frac{3}{4\pi}\cos^2\theta=|Y_{10}|^2\EEQ
Moreover, it holds that
\BEQ \int_0^{2\pi}\frac{\d\nu}{2\pi}\,\hat L_z \,\cY_{1m}^{1m}\,=\,\frac{3\,m}{8\pi}\sin^2\theta,\qquad(m=-1,0,1)
\EEQ
If we integrate this over all angles $\theta$, $\phi$
and make the identification $L_{z\,op}\leftrightarrow\hat L_z$,
we reproduce the quantum result
\BEQ \overline{\, \hat L_z\, \cY_{1m;1m}\,}\equiv\int\sin\theta\d\theta\d\phi\int\frac{\d\nu}{2\pi}
\hat L_z \cY_{1m}^{1m}=m=\langle Y_{1m}|L_{z\,op}|Y_{1m}\rangle,
\qquad (m=-1,0,1)\EEQ
and also the related results
\BEQ  \overline{\, \hat L_x \,\cY_{1m}^{1m}\,}=0=\langle Y_{1m}|L_{x\,\,op}|Y_{1m}\rangle,\qquad
 \overline{\, \hat L_y\, \cY_{1m}^{1m}\,}=0=\langle Y_{1m}|L_{y\,\,op}|Y_{1m}\rangle.
\EEQ

We conclude that at least the lowest non-trivial square of spherical harmonics can be properly generalized.

\renewcommand{\thesection}{\arabic{section}}
\section{Comparison with the Wigner function}
\setcounter{equation}{0}\setcounter{figure}{0} \setcounter{section}{5}
\renewcommand{\thesection}{\arabic{section}.}

The Wigner function of the ground state wavefunction reads in the non-relativistic limit
~\cite{Schleich,Wignerfion}
\BEQ && W(\vr,\vp)=\int\d^3 q\,\psi^\ast(\vp+\half\vq)\psi(\vp-\half\vq)e^{-i\vr\cdot \vq} =
\frac{2^6\pi }{(2\pi)^6}\int\d^3 q\,\frac{e^{-i\vr\cdot\vq}}
{[1+(\vp+\half\vq)^2]^2[1+(\vp-\half\vq)^2]^2}
\EEQ
As opposed to our phase space density, this function is partly negative,
a fact commonly considered as a key argument against
any underlying quasi-classical reality. Its marginal reads
\BEQ \label{margWig} W(\vp)= \frac{8}{\pi^2(1+p^2)^4} \EEQ
The non-relativistic limit of our phase space density (\ref{Pc=})
reads for the ground state
\BEQ \cP(\vr,\vp)=\frac{2}{\pi^3}LR^3e^{-2R} \EEQ
Taking spherical coordinates along $\vp$ yields $L=pr\sin\theta$ and
using $\half p^2-1/r=-1/R$ one derives for its marginal
\BEQ \label{margPSD}
\cP(\vp)=\frac{2p}{\pi}\int_0^{2/p^2}d r\,r^3R^3e^{-2R}
=\frac{2p}{\pi}\int_0^\infty\d R\,\frac{R^6}{(1+\half p^2R)^5}e^{-2R}
\EEQ
Whereas (\ref{margWig}) is finite at $p=0$ and decays as $p^{-8}$ for large $p$,
Eq. (\ref{margPSD}) is linear at $p=0$ and decays as $p^{-9}$ for large $p$.
Figure 5.1 compares these results. It is seen that the difference is mainly at small $p$,
where because of the phase space volume little weight is located.
The tails are already small when they start to deviate from each other.

\begin{figure}
  {\includegraphics[height=.2\textheight]{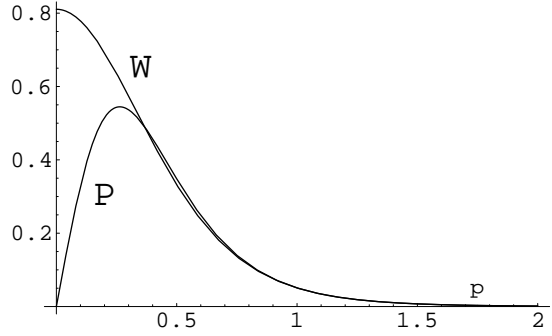}}
  \caption{
Reduced phase space density $\cP(p)$ of the non-relativistic hydrogen ground state,
Eq. (\ref{margPSD}), (lower curve), and reduced  Wigner density $W(p)$, Eq. (\ref{margWig}),
(upper curve),  as function of the momentum $p$, in atomic units. They are both normalized to unity. }
\end{figure}

\section{Discussion}

In this paper we have considered the question whether the quantum probability densities
(the squares of quantum wavefunctions) of the relativistic hydrogen atom,
in the absence of electron spin, may arise from the phase space distribution in an
underlying classical stochastic theory (``sub-quantum theory'' or ``hidden variable theory'').
Because of the small value of the fine structure constant, this is a weak damping problem,
for which it is generally known that the stationary distribution is determined by conserved quantities
(energy and angular momentum) alone. Without specifying the underlying stochastic theory,
we have shown by construction that such a scheme can work for the considered cases.
The relativistic version of the hydrogen problem has been chosen in order to have the fine structure
constant as a parameter, and the generalization works to all orders (be it that from the outset,
all Lamb shift effects are neglected, in our approach as well as in Schr\"odinger dynamics).

Our resulting phase space density $\cP(\vr,\vp)$ does not coincide with the Wigner function $W(\vr,\vp)$.
For the hydrogen ground state the latter can be expressed as derivatives of an integral
~\cite{Schleich, Wignerfion}.
In contrast, for our ground state $\cP(\vr,\vp)$ is explicit, it  contains powers and an exponential.
Although we did not prove its uniquness, it looks simple and has been generalized for the
whole family of Yrast states.
It is well known that the Wigner function for the hydrogen ground state
is partly negative, and this has long been considered
as a no-go argument against any classical-type subquantum structure. However, our $\cP(\vr,\vp)$ is
non-negative for the ground state. Does this solve that paradox? Yes, because our momentum
stands for the instantaneous momentum of the particle. The momentum of the Wigner function
is defined in via off-diagonal elements of the quantum density matrix, and should be
considered as a coarse grained momentum, just as would happen in brownian motion
~\cite{Nelson, AKNbrownentanglement}. For such a coarse grained momentum positivity
of the density is not obvious, probably not needed, and certainly absent.

From a physical point of view, there are many reasons to accept that particles are to good approximation
point particles: traces in cloud chambers, images of non-flat crystal surfaces
obtained by scanning tunnel microscopy, trapping of ions in magnetic fields over periods exceeding a year.
If, one the subquantum level they just have a
position and a speed, the phase space density cannot be negative.
The problem of negative probabilities is solved for our ground state, but still a potential danger
for excited states.
For the $2p$ state the generalized spherical harmonics are partly negative, while for the
$2s$ state the radial phase space density is partly negative.
But in the theory of classical brownian motion the excited states are also partly negative,
while the total probability density will always be non-negative.

Within Stochastic Electrodynamics the $2s$ state can be populated
by a two-photon excitation from an ensemble of ground state atoms,
but it cannot happen in such a manner that the total phase space probability is negative anywhere.
For instance, one could imagine a large population of the $1s$ state and a partial one of the $2s$ state,
with not-too-large values of $L$ and $R$ in the $2s$ state,
such that the total probability is still positive everywhere.
Afterall, at the classical level each realization consists of an electron in some orbit around the nucleus
in the presence of fluctuations, and a ``state'' merely describes an ensemble of such systems.
The $2p$ phase space densities at $m$ fixed are partly negative,
so they cannot exist on their own. However, the `transversally' polarized mixture
$\cP_{21}(E,L)(\cY_{1-1}^{1-1}+\cY_{11}^{11})$
is non-negative, while the unpolarized mixture
$\cP_{21}(E,L)(\cY_{1-1}^{1-1}+\cY_{10}^{10}+\cY_{11}^{11})=3\cP_{21}(E,L)/(4\pi)$
is strictly positive. They can in principle exist without ground state buffer.
The odd parts may differ from the assumed linear ones of (\ref{Lzlinear}).
When they are  cubic, both  $\cY_{1-1}^{1-1}$ and $\cY_{11}^{11}$
can be positive, but not $\cY_{10}^{10}$. The quantum operator $L_{z\,op}$
may then be identified with the classical $\hat L_z$ times a constant.

Given the phase space distribution, we have considered the average energy. We have observed that the
quantum energy eigenvalues are reproduced provided certain correlations are neglected, after which
averages of products
reduce to products of averages. Physically this occurs when these averages are taken at very different times,
such that their fluctuations are uncorrelated. This hints at the possibility that the operator structure of
quantum theory arises from the physics of timescale separated observables in an underlying stochastic theory.
The property that repeatedly applying a quantum operator on one of its eigenstates produces just powers
of the related eigenvalue has a natural interpretation in terms of  temporal separation of averages, as
is  long understood in the theory Stochastic Electrodynamics~\cite{delaPenaCettoBook}.
It is remarkable that such a separation
appears necessary here to establish the proper mapping of energies between the class of stochastic
theories and quantum mechanics.

In atomic physics spectral emissions are due to transition from an excited state to a lower one.
The $2p$-state has a lifetime given by $\alpha^{-3}\tau_0$ multiplied by phase space factors,
the final result is of order of magnitude of the inverse Lamb shift frequency $\sim 1\,ns$.
So about $\alpha^{-3}$ Keppler orbits are traversed before the emission takes place.
This large number of revolutions, together with the gradual influence of the noise that
slowly changes the orbits, may explain the sharpness of spectral lines
within our picture. If only a few orbits are traversed, the sharpness of the
spectral line will not be achieved, since the spread in energies of Keppler orbits is large.
But this just provides a physical picture for
the energy-time uncertainty relation know from quantum mechanics.

There are many remaining and often worrying issues.
An essential item is the role of dynamics, such as the $1s$-$2s$ and $1s$-$2p$ interference terms.
Likewise, there is the task to include spin.
It is good to realize that our present radial distributions are functions of $\vL^2$,
which remains conserved in the presence of spin.

For scattering and ionization experiments on of e.g. electrons on e.g. atomic hydrogen
~\cite{Mott} it is known that the energy loss of the incoming electron can only take one of
the discrete values needed to bring the ground state electron to the
one of the excited levels (Lyman lines).
In the recent work~\cite{Childersetal} transitions to $n=2$, $3$ and
$4$ are clearly observed, while the higher states form a quasi continuum, that matches
the ionization continuum.
Whether these discrete features can be described in the present
formalism is a question related to the existence of excited states,
and the timescale on which they express themselves.
What we have discussed is a number of features that should be respected in our class of
the underlying theories. Whether the proposed type of ground state and excited states exists
and what happens during electron scattering events,
are questions that cannot be answered before specializing the precise type of stochastic noise.

Like any new theory, the present approach poses more questions than it solves.
For any sub-quantum theory a vast amount of experimental constraints is to be respected.
But we are convinced that no individual quantum experiment is described by any current theory.
So if Einstein was right after all, quantum mechanics is incomplete and then there
should be some path to an underlying, more complete theory.
We hope that the present work plays a role in uncovering it.

\section*{Acknowledgments}
It is a pleasure to thank Armen Allahverdyan for discussions that throughout the years
helped to shape this research and Roger Balian for several more.
They, for sure, carry no responsibility for the presented material  and interpretation.

\end{document}